\documentclass[twoside,fleqn, 11pt]{article}

\usepackage{amsmath,amssymb,amsfonts,amsthm,graphics,graphicx,color}
\usepackage[MeX]{polski}
\usepackage[portuguese,spanish,english]{babel}
\usepackage{anysize}
\usepackage{fancyhdr}
\usepackage{easybmat, bm}
\usepackage[mathlines]{lineno}
\usepackage[T1]{fontenc}
\usepackage[useregional]{datetime2}

\usepackage[small, center, skip=2pt]{caption}
\DeclareCaptionLabelFormat{nospace}{#1#2}

\usepackage{titlesec}
\titleformat{\section}
  {\bfseries  \fontsize{11}{11} \centering} 
  {\thesection.}{0.5em}{}

\titleformat{\subsection}
  {\fontsize{11}{11}\bfseries\sffamily}
  {\thesubsection.}{0.2em}{}
  
\titleformat{\subsubsection}
  {\fontsize{11}{11}\bfseries\sffamily}
  {\subsubsection.}{0.2em}{}  
  
\newenvironment{BLtitle}
    {\begin{center}
    \vspace*{1.2cm}
    \large \bf
    }
    {
    \end{center}
    }  
    
\newenvironment{BLauthor}
    {\begin{center}
    \bf
    }
    {
    \end{center}
    }     
\newenvironment{BLaffiliation}
    {\begin{center}
    \footnotesize
    \vspace{0.2cm}
    }
    {
    \end{center}
  \vspace{0.2cm}
    }

\newenvironment{Summary}
{\begin{minipage}{11.4cm}
\thispagestyle{empty}
\begin{center}
\textsc{Summary\vspace{.2cm}}
\end{center}
\small
}
{
\end{minipage} \\[.2cm]   
}

\newenvironment{Key_words}
{\begin{minipage}{11.4cm}
\thispagestyle{empty}
\small
\textbf{Key words:}
}
{
\end{minipage}    
}

\newenvironment{BLReferences}
{\begin{center}
\textsc{References\vspace{-1.2cm}}
\end{center}

\small
\begin{thebibliography}{xx}
}
{
\end{thebibliography}
}
\usepackage{natbib}
\usepackage{graphicx}

\theoremstyle{plain}
\newtheorem{thm}{Proposition}

\theoremstyle{definition}

\newcommand{\h}{\mathcal{H}}

\DeclareMathOperator{\BIC}{BIC}
\DeclareMathOperator{\BF}{BF}

\begin{document}
\vspace*{-4.0cm}
\begin{flushright}
\begin{figure}[h!]
\begin{flushleft}
\includegraphics[scale=0.15]{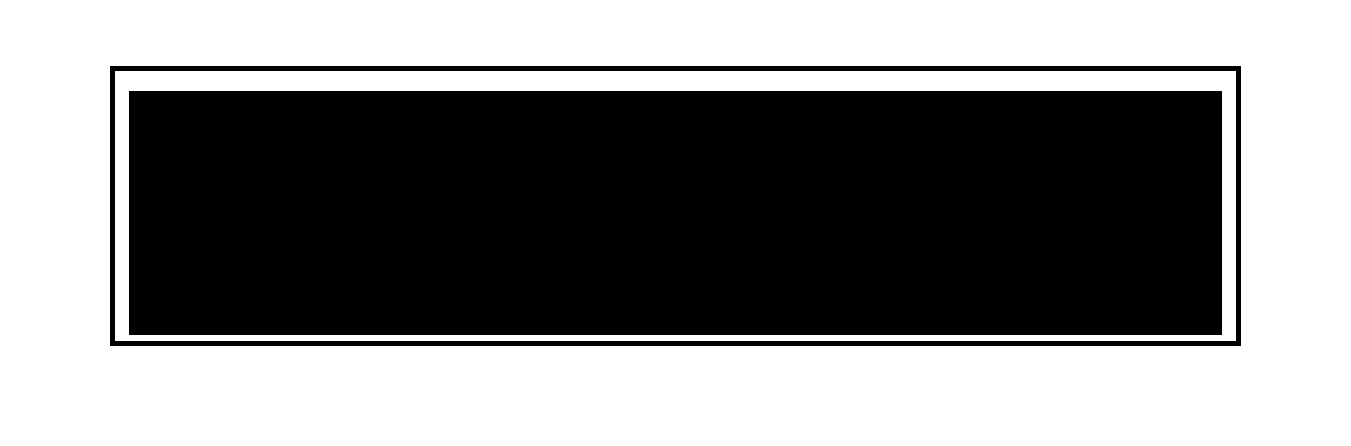}
\end{flushleft}
\end{figure}
\vspace*{-1.2cm}
\footnotesize\textit{\textcolor{black}{DOI: xx.xxxx/xxx-xxx-xxxx } }
\end{flushright}
\begin{center}
\vspace*{0.46cm}
\small \footnotesize
\itshape In press at Biometrical Letters\\[-.01cm]
\today
\end{center} 

\begin{BLtitle}
Computing analytic Bayes factors from summary statistics in repeated-measures designs
\end{BLtitle}


\begin{BLauthor}
Thomas J. Faulkenberry$^{1}$ and Keelyn B. Brennan$^{1}$
\end{BLauthor}


%
\begin{BLaffiliation}
$^1$Department of Psychological Sciences, Tarleton State University, Stephenville, Texas, 76402, USA, e-mail: faulkenberry@tarleton.edu \\
\end{BLaffiliation}

%
%
\begin{Summary}
 Bayes factors are an increasingly popular tool for indexing evidence from experiments. For two competing population models, the Bayes factor reflects the relative likelihood of observing some data under one model compared to the other. Computing Bayes factors can be difficult, requiring integrating the product of the likelihood and a prior distribution on the population parameter(s) for both competing models. Previous work has obviated this difficulty for independent-groups designs. In this paper, we develop a new analytic formula for computing Bayes factors directly from minimal summary statistics in repeated-measures designs. This work is an improvement on previous methods for computing Bayes factors from summary statistics (e.g., the BIC method), which produce Bayes factors that violate the Sellke upper bound of evidence for smaller sample sizes. The new approach taken in this paper extends requires knowing only the $F$-statistic and degrees of freedom, both of which are commonly reported in most empirical work. In addition to providing computational examples, we report a simulation study that benchmarks the new formula against other methods for computing Bayes factors in repeated-measures designs. Our new method provides an easy way for researchers to compute Bayes factors directly from a minimal set of summary statistics, allowing users to index the evidential value of their own data, as well as data reported in published studies.
\end{Summary}
%

\begin{Key_words}
Bayes factor; Pearson Type VI distribution; summary statistics; repeated-measures.
\end{Key_words}

\setcounter{page}{1}


\section{Introduction}

In this paper, we develop an analytic formula to compute Bayes factors for repeated-measures designs using only minimal summary statistics from the analysis of variance. Previous attempts to quantify evidence from summary statistics in repeated-measures designs have all relied upon the BIC approximation \citep[e.g.,][]{wagenmakers2007,faulkenberry2018,faulkenberry2020}. In contrast, our new method avoids the need for approximation and produces an exact (analytic) Bayes factor directly from the observed $F$-statistic and associated degrees of freedom. This paper extends the development of the between-subjects {\it Pearson Bayes factor} \citep{faulkenberry2021} to also consider repeated-measures designs, thus widening the scope and its use for indexing evidential value from summary statistics. Further, our formula is gives researchers a way to compute Bayes factors in repeated-measures designs that does not involve integration \citep{rouder2012} or approximation \citep{nathoo2016,faulkenberry2020}. As such, users can easily compute Bayes factors for their own repeated-measures data, and also for any results that are reported in the scientific literature -- even null effects. Our method thus further affords researchers the ability to gauge the evidential value of a collection of published results in a straightforward way without the need for raw data.

\section{Background}

We begin with some background on inference in experimental designs with repeated measurements. Let us consider an experiment where $k$ repeated measurements are recorded for each of $n$ experimental subjects, giving a collection of $N=nk$ observations $\bm{y}$. We then place a linear mixed-effects model on the collection of observations:%
\[
  y_{ij} = \mu + \alpha_j + \pi_i + \varepsilon_{ij}; \hspace{3mm} i=1,\dots,n;\hspace{1mm}j=1\cdots,k,
\]%
where $\mu$ represents the grand mean, $\alpha_j$ represents the treatment effect associated with group $j$, $\pi_i$ represents the effect of subject $i$, and $\varepsilon_{ij} \sim \mathcal{N}(0,\sigma_{\varepsilon}^2)$. Note that the repeated-measures design induces a correlated structure on the data, so the $N=nk$ observations are not independent. As a result, we have $n(k-1)$ independent observations \citep{masson2011}.

Ultimately, we want to know whether there are differences among the treatment groups induced by the repeated measurements. We can answer this by applying a hypothesis test, where we consider the predictive ulility of two competing models:%
\begin{gather*}
  \mathcal{H}_0:\alpha_j=0\text{ for }j=1,\dots,k\\
  \mathcal{H}_1:\alpha_j\neq 0\text{ for some }j
\end{gather*}

A classical approach to model selection in this context is the analysis of variance (ANOVA) procedure \citep{fisher1925}, which proceeds by partitioning the total variance $SST$ in the data $\bm{y}$ into two sources: $SSA$, which denotes the variance between the treatment groups, and $SSR$, which denotes the residual variance left over after accounting for treatment variability. The $F$-statistic for the treatment effect is computed as%
\[
  F=\frac{SSA}{SSR}\cdot \frac{df_{\text{residual}}}{df_{\text{treatment}}} = \frac{SSA}{SSR}\cdot \frac{(n-1)(k-1)}{k-1} = \frac{SSA}{SSR}\cdot (n-1).
    \]%
As a proxy for our observed data, the $F$ statistic can be used to assess model fit by transforming it to a $p$-value, which represents the likelihood of the observed data $\bm{y}$ under the null hypothesis $\mathcal{H}_0$. If the $p$-value is small, we conclude that the data are unlikely to have occurred under $\mathcal{H}_0$ and thus reject $\mathcal{H}_0$ in favor of the alternative hypothesis $\mathcal{H}_1$.

Unfortunately, this classical approach to inference is fraught with some issues that undermine its use \citep[e.g.,][]{wagenmakers2007}. In light of this, we advocate a Bayesian approach to model selection instead and use Bayes factors \citep{kassRaftery1995}. The Bayes factor $\text{BF}_{01}$ is defined as the ratio of marginal likelihoods for $\mathcal{H}_0$ and $\mathcal{H}_1$, respectively. That is,%
\begin{equation}\label{eq:marginal}
 \text{BF}_{01} = \frac{p(\bm{y}\mid \mathcal{H}_0)}{p(\bm{y}\mid \mathcal{H}_1)}.
\end{equation}%
This ratio indexes the relative likelihood of observing data $\bm{y}$ under $\mathcal{H}_0$ compared to $\mathcal{H}_1$, so $\text{BF}_{01}>1$ is taken as evidence for $\mathcal{H}_0$ over $\mathcal{H}_1$. Similarly, $\text{BF}_{01}<1$ is taken as evidence for $\mathcal{H}_1$. Note that $\text{BF}_{01} = 1/\text{BF}_{10}$, so if $\text{BF}_{01}<1$, we can take the reciprocal and equivalently write $BF_{10} > 1$. For this reason, it is typical to report a Bayes factor as a number greater than 1. For example, instead of reporting $\text{BF}_{01} = 0.25$, we will take the reciprocal and report $\text{BF}_{10} = 1/0.25 = 4$. Both representations imply that the observed data are 4 times more likely under $\mathcal{H}_1$ than under $\mathcal{H}_0$.

Unlike $p$-values, Bayes factors can be directly transformed into posterior probabilities \citep{faulkenberry2021}:%
\[
  p(\mathcal{H}_0\mid \bm{y}) = \frac{\text{BF}_{01}\cdot p(\mathcal{H}_0)}{\text{BF}_{01}\cdot p(\mathcal{H}_0) + p(\mathcal{H}_1)}
\]%
and%
\[
  p(\mathcal{H}_1\mid \bm{y}) = \frac{\text{BF}_{10}\cdot p(\mathcal{H}_1)}{\text{BF}_{10}\cdot p(\mathcal{H}_1) + p(\mathcal{H}_0)}.
\]%
Commonly, we take a default assumption that both models are equally likely {\it a priori}, permitting us to set $p(\mathcal{H}_0)=p(\mathcal{H}_1) = 0.5$. In this case, we get the following simplified formulas:%
\begin{equation}\label{eq:probability}
  p(\mathcal{H}_0\mid \bm{y}) = \frac{\text{BF}_{01}}{\text{BF}_{01}+1}, \hspace{1cm} p(\mathcal{H}_1\mid \bm{y}) = \frac{\text{BF}_{10}}{\text{BF}_{10}+1}.
\end{equation}

Much work on Bayes factors during the last 30 years has focused on developing methods to compute $\text{BF}_{01}$ in various designs. Early approaches were based on using approximations of the marginal likelihoods. One well-known example of this approach is called the BIC approximation \citep{raftery1995, kassRaftery1995, wagenmakers2007, masson2011}. The first step of this method is to compute the Bayesian information criterion (BIC) of \citet{schwarz1978} for each model $\h_i$:%
\begin{equation}\label{eq:bic2}
  \BIC(\h_i)=-2\ln L_i + k_i\ln N,
\end{equation}%
where $L_i$ is the maximum likelihood estimate for model $\h_i$, $k_i$ is the number of parameters in $\h_i$, and $N$ is the total number of independent observations in $\bm{y}$. Then, the Bayes factor may be approximated as
\begin{equation}\label{eq:bicBF}
  \BF_{01} \approx \exp \Biggl(\frac{\BIC(\h_1) - \BIC(\h_0)}{2}\Biggr).
\end{equation}

To use Equation \ref{eq:bicBF} for computing Bayes factors, we need only know the BIC values for models $\h_0$ and $\h_1$. In the context of analysis of variance presented earlier, the BIC can be calculated \citep{raftery1995} as%
\[
BIC = N\ln \Biggl(\frac{SSR}{SST}\Biggr) + k\ln N ; .
\]

One downside to the BIC method is that it requires users to have the ``raw'' data available in order to compute $SSR$ and $SST$. \citet{faulkenberry2018} improved Equation \ref{eq:bicBF} for between-subjects designs by recasting the Bayes factor computation to a form that requires only summary statistics. \citet{faulkenberry2020} then extended that result to the context of repeated-measures designs by deriving the formula: 

\begin{equation}\label{eq:bic2}
          \text{BF}_{01} \approx \sqrt{(nk-n)^{k-1}\cdot \Biggl(1+\frac{F}{n-1}\Biggr)^{n-nk}} \; .
\end{equation}%

\begin{figure}
  \centering
  \includegraphics[width=0.7\textwidth]{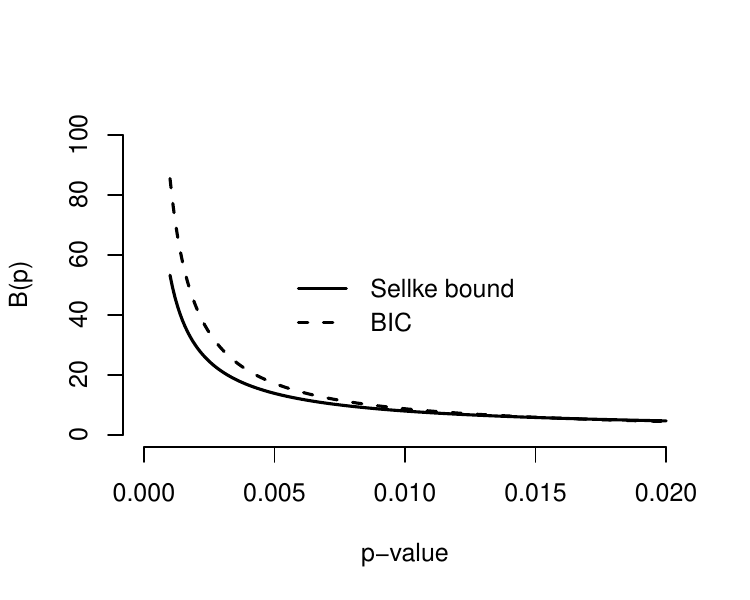}
  \caption{Plot showing that the repeated-measures BIC Bayes factor (dashed line) of \citet{faulkenberry2020} is greater than the Sellke bound (solid line) for $p$-values ranging between $p\approx 0$ and $p=0.02$. }
  \label{fig:sellke}
\end{figure}

The following example illustrates the use of (and a problem with) Equation \ref{eq:bic2}. In a repeated-measures ($k=2$) study with $n=18$ participants, \citet{fayol2012} observed that mean solution times to subtraction problems were 43 milliseconds faster when the subtraction sign was briefly presented before the actual problem itself, $F(1,17) = 27.17$, $p<0.001$. Using Equation \ref{eq:bic2}, we can compute the BIC Bayes factor for these observed data as:%
\begin{align*}
  \text{BF}_{01} & \approx \sqrt{(nk-n)^{k-1}\cdot \Biggl(1+\frac{F}{n-1}\Biggr)^{n-nk}}\\
                 & = \sqrt{(18\cdot 2-18)^{2-1}\cdot \Biggl(1+\frac{27.17}{18-1}\Biggr)^{18-18\cdot 2}}\\
                 & = \sqrt{18^1 \cdot \Biggl(1 + \frac{27.17}{17}\Biggr)^{-18}}\\
                 & = \sqrt{18\cdot (2.5982)^{-18}}\\
                 & = 0.0007863 \;.
\end{align*}%
By taking the reciprocal and casting this Bayes factor as support for $\mathcal{H}_1$, we have $\text{BF}_{10} \approx 1/0.0007863 = 1271.79$. Thus, the BIC Bayes factor tells us that the observed data are approximately 1272 times more likely under $\mathcal{H}_1$ than under $\mathcal{H}_0$. But recall that the Bayes factor derived from Equation \ref{eq:bicBF} is an {\it approximation}. \citet{sellke2001} provided an upper bound for the Bayes factor that can be computed directly from the $p$-value:%
\[
  \BF_{10} \leq -\frac{1}{e\cdot p\ln(p)}.
\]%
Substituting $p=0.0000704$ from our example gives the upper bound%
\begin{align*}
  \BF_{10} & \leq -\frac{1}{e\cdot 0.0000704\cdot \ln(0.0000704)}\\
          &=546.53.
\end{align*}%
From this, the limitation of the BIC approximation is clear. Our computed Bayes factor of 1272 greatly exceeds the Sellke bound of 546.53. In fact, Figure \ref{fig:sellke} shows that this problem persists over a large range of $p$-values. In the figure, the solid line represents the Sellke bound $B(p)$ for $p$-values ranging between 0 and 0.02. The dashed line represents the associated repeated-measures BIC Bayes factor of \citet{faulkenberry2020}, given design parameters equivalent to \citet{fayol2012} (i.e., $k=2$ repeated-measures conditions and $n=18$ subjects).

\section{Analytic Bayes factors for repeated-measures designs}

Against the background of the previous section, we think the need for an analytic Bayes factor for repeated-measures designs is well motivated. Fortunately, recent work by \citet{wangSun} takes an important first step toward this goal. Wang and Sun, motivated by the the work of \citet{gds}, started with a random effects linear model on the observed data $\bm{y}$:%
\[
  y_{ij}=\mu + \alpha_j + \pi_i + \varepsilon_{ij},
\]%
where $\alpha_j \sim \mathcal{N}(0,\sigma^2_a)$, $\pi_i \sim \mathcal{N}(0,\sigma^2_p)$ and $\varepsilon_{ij} \sim \mathcal{N}(0,\sigma^2)$. Cast in this slightly different context of random effects versus the classical fixed effects model\footnote{For one-way analysis of variance, the Bayes factor is equivalent for fixed-effects and random-effects designs \citep{rouder2012}}, the analysis of variance procedure amounts to testing whether the random effects term $\alpha_j$ is identically 0. To capture this constraint, the competing models are defined in a slightly different manner:%
\[
  \h_0:\sigma^2_a=0 \text{ versus }\h_1:\sigma_a^2\neq 0.
  \]

\begin{figure}[h]
  \centering
  \includegraphics[width=0.8\textwidth]{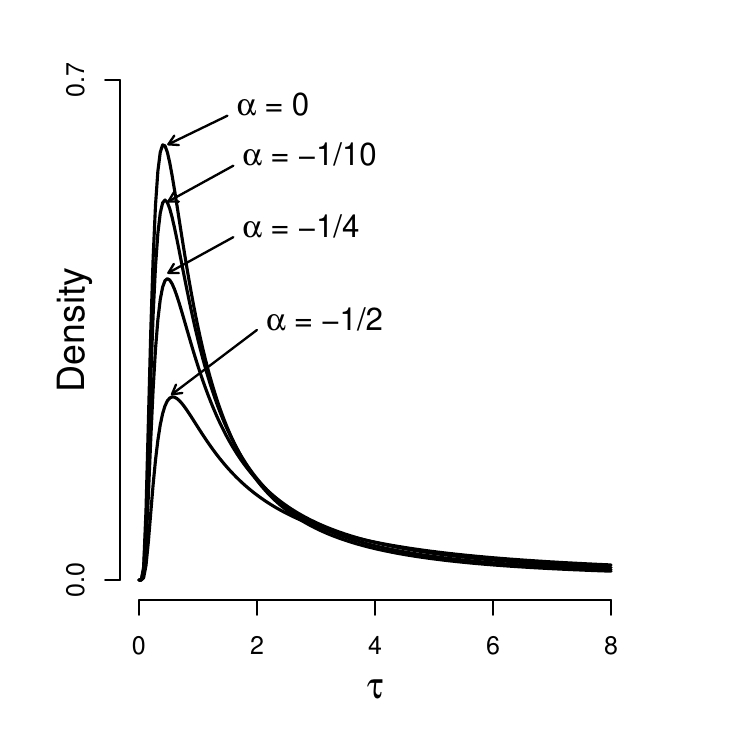}
  \caption{A Pearson Type VI prior for $\tau$, plotted as a function of shape parameter $\alpha$. In this example, we take $n=18$ and $k=2$, but the general pattern holds for other values of $n$ and $k$.}
  \label{fig:prior}
\end{figure}

With this setup, \citet{gds} placed noninformative priors on $\mu$ and $\sigma$ under both $\h_0$ and $\h_1$ and considered a proper prior on the ratio of variance components $\tau = \sigma^2_a/\sigma^2$ under $\h_1$. With this prior specification, Garcia-Donato and Sun showed%
\begin{equation}
  \label{eq:gds}
  BF_{10} = \int_0^{\infty}(1+\tau n)^{\frac{1-k}{2}}\Biggl(1-\frac{\tau n}{1+\tau n}\cdot \frac{SSA}{SST}\Biggr)^{\frac{1-N}{2}}\cdot \pi(\tau)d\tau
\end{equation}%
where $\pi(\tau)$ is left up to the analyst to choose. \citet{wangSun} used a Pearson Type VI distribution, given by:%
\[
  \pi^{PT}(\tau) = \frac{\kappa(\kappa \tau)^{\beta}(1+\kappa \tau)^{-\alpha-\beta-2}}{\mathcal{B}(\alpha+1, \beta+1)}I_{(0,\infty)}(\tau)
\]%
where  $\alpha>-1$ and $\beta>-1$ are shape parameters and $\kappa>0$ is a scale parameter, and $\mathcal{B}(x,y) = \int_0^1t^{x-1}(1-t)^{y-1}dt$ is the standard Beta function. Wang and Sun further reduced the problem of specifying the prior $\pi^{PT}$ to choosing one single parameter $\alpha \in [-\frac{1}{2}, 0]$. They did this by taking $\kappa=n$ and $\beta = \frac{N-k}{2}-\alpha-2$. A plot of $\pi^{PT}$ can be seen in Figure \ref{fig:prior}; here, we take the values $n=18$, $k=2$, and $N=36$ from our example above, thus setting $\kappa=18$ and $\beta=\frac{36-2}{2}-\alpha-2$, where $\alpha$ ranges among the values $-\frac{1}{2}, -\frac{1}{4}, -\frac{1}{10}, 0$.  As we can see, as $\alpha$ decreases from to 0 to $-\frac{1}{2}$, $\tau$ becomes more dispersed and less peaked around the mode. This places more prior mass on larger treatment effects than we would see for values of $\alpha$ closer to 0. Note that this figure is specific to the experimental design (i.e., $n=18$ subjects completing $k=2$ repeated measures conditions), but the described pattern holds for other design parameters.

Given this choice of prior and simplified parameterization, \citet{wangSun} proved that the Bayes factor derived by Garcia-Donato and Sun (Equation \ref{eq:gds}) simplifies to an {\it analytic} expression without the need for integral representation:%
\begin{equation}\label{eq:ws}
 \BF_{10} = \frac{\Gamma\Bigl(\frac{k}{2}+\alpha + \frac{1}{2}\Bigr)\cdot \Gamma\Bigl(\frac{N-k}{2}\Bigr)}{\Gamma\Bigl(\frac{N-1}{2}\Bigr)\cdot \Gamma (\alpha+1)} \cdot \Biggl(\frac{SSR}{SST}\Biggr)^{\alpha-\frac{N-k-2}{2}} \; .
\end{equation}%

To extend the Wang and Sun method for computing $\text{BF}_{10}$ (Equation \ref{eq:ws}) to a repeated-measures context, we will apply a method of \citet{masson2011} and replace $N$ by $n(k-1)$ \citep[see also ][]{campbell2012,bortolussi2002,faulkenberry2020}. This direct substitution readily gives:%
\begin{align} \label{eq:rm1}
  \text{BF}_{10} &= \frac{\Gamma\Bigl(\frac{k}{2}+\alpha + \frac{1}{2}\Bigr)\cdot \Gamma\Bigl(\frac{n(k-1)-k}{2}\Bigr)}{\Gamma\Bigl(\frac{n(k-1)-1}{2}\Bigr)\cdot \Gamma (\alpha+1)} \cdot \Biggl(\frac{SSR}{SST}\Biggr)^{\alpha-\frac{n(k-1)-k-2}{2}} \nonumber \\
  &= \frac{\Gamma\Bigl(\frac{k}{2}+\alpha + \frac{1}{2}\Bigr)\cdot \Gamma\Bigl(\frac{nk-n-k}{2}\Bigr)}{\Gamma\Bigl(\frac{nk-n-1}{2}\Bigr)\cdot \Gamma (\alpha+1)} \cdot \Biggl(\frac{SSR}{SST}\Biggr)^{\alpha-\frac{nk-n-k-2}{2}} \; .
\end{align}

We now consider some identities that will greatly simplify Equation \ref{eq:rm1}. First, let $x$ denote the between-groups degrees of freedom, $df_{\text{treatment}}$, and let $y$ denote the residual degrees of freedom, $df_{\text{residual}}$. As seen earlier, this gives $x=k-1$ and $y=(n-1)(k-1) = nk - n - k +1$. From here, we can derive the following three identities:%
\[
  k=x+1 \; ;
\]

\begin{align*}
        nk-n-k &= (nk-n-k+1)-1\\
               &= y-1 \; ;
\end{align*}

\begin{align*}
        nk-n-1 &= (nk-n-k+1)+k-2\\
               &= y+(x+1)-2\\
               &= x + y - 1 \; .
\end{align*}

\noindent
We now substitute these identities into Equation \ref{eq:rm1}, giving:%
\begin{align}\label{eq:rm2}
  \text{BF}_{10} &= \frac{\Gamma\Bigl(\frac{k}{2}+\alpha + \frac{1}{2}\Bigr)\cdot \Gamma\Bigl(\frac{nk-n-k}{2}\Bigr)}{\Gamma\Bigl(\frac{nk-n-1}{2}\Bigr)\cdot \Gamma (\alpha+1)} \cdot \Biggl(\frac{SSR}{SST}\Biggr)^{\alpha-\frac{nk-n-k-2}{2}} \nonumber \\
                 &= \frac{\Gamma\Bigl(\frac{x+1}{2}+\alpha + \frac{1}{2}\Bigr)\cdot \Gamma\Bigl(\frac{y-1}{2}\Bigr)}{\Gamma\Bigl(\frac{x+y-1}{2}\Bigr)\cdot \Gamma (\alpha+1)} \cdot \Biggl(\frac{SSR}{SST}\Biggr)^{\alpha-\frac{(y-1)-2}{2}}\; .
\end{align}%
By definition, we have
\[
  F =\frac{SSA}{SSR}\cdot \frac{df_{\text{residual}}}{df_{\text{treatment}}} = \frac{SSA}{SSR}\cdot \frac{y}{x} \; .
\]%
Also, in a repeated-measures design, subject variability is removed from the total sum of squares term $SST$. Thus, we can replace $SST$ with $SSA+SSR$, giving%
\begin{align*}
  \frac{SST}{SSR} &= \frac{SSA+SSR}{SSR}\\
                  &= \frac{SSA}{SSR}+1\\
                  &= \frac{xF}{y} + 1\\
                  &=\frac{y+xF}{y} \; .
\end{align*}%
Taking the reciprocal of this identity and subsituting back into Equation \ref{eq:rm2} proves the following proposition:%

\begin{thm}\label{mainTheorem}
  Given a repeated-measures analysis of variance summary reported in the form $F(x,y)$, where $x$ equals the between-treatments degrees of freedom and $y$ equals the residual degrees of freedom, the Bayes factor can be expressed in analytic form as%
  \[
    \text{BF}_{10} = \frac{\Gamma\Bigl(\frac{x}{2}+\alpha + 1\Bigr)\cdot \Gamma\Bigl(\frac{y-1}{2}\Bigr)}{\Gamma\Bigl(\frac{x+y-1}{2}\Bigr)\cdot \Gamma (\alpha+1)} \cdot \Biggl(\frac{y}{y+xF}\Biggr)^{\alpha-\frac{y-3}{2}} \; ,
  \]

 where $\alpha \in \Bigl[-\frac{1}{2},0\Bigr]$.
\end{thm}

\section{Example computations}

In this section, we provide two examples of computation of the repeated-measures analytic Bayes factor given in Proposition \ref{mainTheorem}. Both examples come from \citet{fayol2012}, which was briefly mentioned in the previous section on BIC Bayes factors. In their study, \citet{fayol2012} measured adults' response times on a computerized single-digit mental arithmetic task. On some problems, the arithmetic operator (e.g., the addition sign or the multiplication sign) appeared on a computer screen 150 milliseconds before the operands, whereas on other problems, the operator and operands appeared simultaneously. Two critical results appeared and are worth further consideration in our example computations. First, \citet{fayol2012} observed that addition problems for which the operator appeared 150 milliseconds before the operands were solved significantly faster than those for which the operator and operands appeared simultaneously, $F(1,17) = 52.36$, $p<0.001$. Second, they observed that this pattern did {\it not} occur on multiplication problems, $F(1,17) = 1.75$, $p=0.20$. On the basis of this claimed null effect, Fayol and Thevenot reasoned that mental processes for addition must be fundamentally different from those involved in multiplication.

We can use Proposition \ref{mainTheorem} to assess the evidence for the claims of \citep{fayol2012}. We begin by defining the following model on the mean response times $y_{ij}$:%
\[
  y_{ij}=\mu + \alpha_j + \pi_i + \varepsilon_{ij},
\]%
where $\mu$ is the grand mean, $\alpha_j \sim \mathcal{N}(0,\sigma^2_a)$, $\pi_i \sim \mathcal{N}(0,\sigma^2_p)$ and $\varepsilon_{ij} \sim \mathcal{N}(0,\sigma^2)$. Critically, $\alpha_j$ ($j=1,2$ for this example) codes the operator preview effect: $\alpha_1$ is the grand mean adjustment for trials in which the operator was displayed simultaneously with the operands, and $\alpha_2$ is the grand mean adjustment for trials in which the operator was displayed 150 milliseconds before the operands. The question of interest is whether the effect term $\alpha_j$ is identically 0, which we assess by determining whether the variance of the random effect term is 0. This leads to the following two competing models:%
\[
  \h_0:\sigma^2_a=0 \text{ versus }\h_1:\sigma_a^2\neq 0.
  \]

We can use our repeated-measures analytic Bayes factor formula from Proposition \ref{mainTheorem} to index the evidence for $\mathcal{H}_1$ and $\mathcal{H}_0$ that come from the addition and multiplication data, respectively. First, let's compute the Bayes factor for the addition result. In addition to the observed $F$-statistic (52.36) and the relevant degrees of freedom ($x=1$, $y=17$), we must also specify $\alpha$, the width of the prior distribution on the variance ratio $\tau$. We will employ a ``bracketing'' approach and compute the Bayes factor at both ends of its consistency range; that is, we will use both $\alpha=-1/2$ and $\alpha=0$. Beginning with $\alpha=-1/2$, we substitute the summary statistics into the formula, obtaining the following:%
\begin{align*}
  \text{BF}_{10} &= \frac{\Gamma\Bigl(\frac{x}{2}+\alpha + 1\Bigr)\cdot \Gamma\Bigl(\frac{y-1}{2}\Bigr)}{\Gamma\Bigl(\frac{x+y-1}{2}\Bigr)\cdot \Gamma (\alpha+1)} \cdot \Biggl(\frac{y}{y+xF}\Biggr)^{\alpha-\frac{y-3}{2}}\\
                 &=\frac{\Gamma\Bigl(\frac{1}{2}-\frac{1}{2}+1\Bigr) \cdot \Gamma\Bigl(\frac{17-1}{2}\Bigr)}{\Gamma\Bigl(\frac{1+17-1}{2}\Bigr)\cdot\Gamma\Bigl(-\frac{1}{2}+1\Bigr)}\Biggl(\frac{17}{17+1\cdot 52.36}\Biggr)^{-\frac{1}{2}-\Bigl(\frac{17-3}{2}\Bigr)}\\
                 &=\frac{\Gamma\bigl(1\bigr) \cdot \Gamma\bigl(8\bigr)}{\Gamma\Bigl(\frac{17}{2} \Bigr) \cdot \Gamma\Bigl(\frac{1}{2}\Bigr)}\Biggl(\frac{17}{69.36}\Biggr)^{-\frac{15}{2}}\\
              &=\frac{1 \cdot 5040}{14034.41 \cdot 1.772454} \bigl( 0.245098 \bigr)^{-7.5}\\
&=7702.17 \; .
\end{align*}%
Using Equation \ref{eq:probability}, we can convert this Bayes factor to a posterior probability. Assuming equal prior odds for $\mathcal{H}_0$ and $\mathcal{H}_1$, the posterior probability for $\mathcal{H}_1$ is:%
\begin{align*}
  p(\mathcal{H}_1\mid \bm{y}) &= \frac{\text{BF}_{10}}{\text{BF}_{10}+1}\\
                              &= \frac{7702.17}{7702.17+1}\\
  &= 0.99987 \;.
\end{align*}

Now, we repeat this calculation, but with $\alpha=0$. This gives the following:%
\begin{align*}
  \text{BF}_{10} &= \frac{\Gamma\Bigl(\frac{x}{2}+\alpha + 1\Bigr)\cdot \Gamma\Bigl(\frac{y-1}{2}\Bigr)}{\Gamma\Bigl(\frac{x+y-1}{2}\Bigr)\cdot \Gamma (\alpha+1)} \cdot \Biggl(\frac{y}{y+xF}\Biggr)^{\alpha-\frac{y-3}{2}}\\
                 &=\frac{\Gamma\Bigl(\frac{1}{2}+0+1\Bigr) \cdot \Gamma\Bigl(\frac{17-1}{2}\Bigr)}{\Gamma\Bigl(\frac{1+17-1}{2}\Bigr)\cdot\Gamma\bigl(0+1\bigr)}\Biggl(\frac{17}{17+1\cdot 52.36}\Biggr)^{0-\Bigl(\frac{17-3}{2}\Bigr)}\\
                 &=\frac{\Gamma\Bigl(\frac{3}{2}\Bigr) \cdot \Gamma\bigl(8\bigr)}{\Gamma\Bigl(\frac{17}{2} \Bigr) \cdot \Gamma\bigl(1 \bigr)}\Biggl(\frac{17}{69.36}\Biggr)^{-\frac{14}{2}}\\
              &=\frac{0.8862269 \cdot 5040}{14034.41 \cdot 1} \bigl( 0.245098 \bigr)^{-7}\\
&=5989.80 \; .
\end{align*}%
Assuming equal prior odds for $\mathcal{H}_0$ and $\mathcal{H}_1$, the posterior probability for $\mathcal{H}_1$ can be computed (using Equation \ref{eq:probability}) as:%
\begin{align*}
  p(\mathcal{H}_1\mid \bm{y}) &= \frac{\text{BF}_{10}}{\text{BF}_{10}+1}\\
                              &= \frac{5989.80}{5989.80+1}\\
  &= 0.99983 \;.
\end{align*}

Thus, the data observed by \citet{fayol2012} are between 5989.80 and 7702.17 times more likely under $\mathcal{H}_1$ than under $\mathcal{H}_0$, with posterior probability for $\mathcal{H}_1$ between 0.99983 and 0.99987. In all, these data give {\it substantial} evidence for an operator priming effect on addition.

What can be said about the evidence for $\mathcal{H}_0$ from their data for multiplication? Recall that they did not find a significant operator priming effect for multiplication, $F(1,17)=1.75$, $p=0.20$. We can now perform a similar computation to gauge this evidence exactly. Proceeding as before with $\alpha=-1/2$, we obtain:%
\begin{align*}
  \text{BF}_{10} &= \frac{\Gamma\Bigl(\frac{x}{2}+\alpha + 1\Bigr)\cdot \Gamma\Bigl(\frac{y-1}{2}\Bigr)}{\Gamma\Bigl(\frac{x+y-1}{2}\Bigr)\cdot \Gamma (\alpha+1)} \cdot \Biggl(\frac{y}{y+xF}\Biggr)^{\alpha-\frac{y-3}{2}}\\
                 &= \frac{\Gamma\Bigl(\frac{1}{2}-\frac{1}{2} + 1\Bigr)\cdot \Gamma\Bigl(\frac{17-1}{2}\Bigr)}{\Gamma\Bigl(\frac{1+17-1}{2}\Bigr)\cdot \Gamma \Bigl(-\frac{1}{2}+1\Bigr)} \cdot \Biggl(\frac{17}{17+1\cdot 1.75}\Biggr)^{-\frac{1}{2}-\frac{17-3}{2}}\\
                 &=\frac{\Gamma\bigl(1\bigr) \cdot \Gamma\bigl(8\bigr)}{\Gamma\Bigl(\frac{17}{2} \Bigr) \cdot \Gamma\Bigl(\frac{1}{2}\Bigr)}\Biggl(\frac{17}{18.75}\Biggr)^{-\frac{15}{2}}\\
                 &=\frac{1 \cdot 5040}{14034.41 \cdot 1.772454} \bigl( 0.90666667 \bigr)^{-7.5}\\
  &=0.4225 \; .
\end{align*}%
Since the obtained Bayes factor is less than 1, we take the reciprocal to cast it as evidence for $\mathcal{H}_0$:%
\begin{align*}
  \text{BF}_{01} &= \frac{1}{\text{BF}_{10}}\\
                 &= \frac{1}{0.4225}\\
  &=2.37 \;.
\end{align*}%
Assuming equal prior odds for $\mathcal{H}_0$ and $\mathcal{H}_1$, the posterior probability for $\mathcal{H}_0$ can be computed via Equation \ref{eq:probability}:%
\begin{align*}
  p(\mathcal{H}_0\mid \bm{y}) &= \frac{\text{BF}_{01}}{\text{BF}_{01}+1}\\
                              &= \frac{2.37}{2.37+1}\\
  &= 0.70326 \;.
\end{align*}

Similarly, we can repeat the computation with $\alpha=0$:%
\begin{align*}
  \text{BF}_{10} &= \frac{\Gamma\Bigl(\frac{x}{2}+\alpha + 1\Bigr)\cdot \Gamma\Bigl(\frac{y-1}{2}\Bigr)}{\Gamma\Bigl(\frac{x+y-1}{2}\Bigr)\cdot \Gamma (\alpha+1)} \cdot \Biggl(\frac{y}{y+xF}\Biggr)^{\alpha-\frac{y-3}{2}}\\
                 &=\frac{\Gamma\Bigl(\frac{1}{2}+0+1\Bigr) \cdot \Gamma\Bigl(\frac{17-1}{2}\Bigr)}{\Gamma\Bigl(\frac{1+17-1}{2}\Bigr)\cdot\Gamma\bigl(0+1\bigr)}\Biggl(\frac{17}{17+1\cdot 1.75}\Biggr)^{0-\Bigl(\frac{17-3}{2}\Bigr)}\\
                 &=\frac{\Gamma\Bigl(\frac{3}{2}\Bigr) \cdot \Gamma\bigl(8\bigr)}{\Gamma\Bigl(\frac{17}{2} \Bigr) \cdot \Gamma\bigl(1 \bigr)}\Biggl(\frac{17}{18.75}\Biggr)^{-\frac{14}{2}}\\
              &=\frac{0.8862269 \cdot 5040}{14034.41 \cdot 1} \bigl( 0.90666667 \bigr)^{-7}\\
&=0.6319 \; .
\end{align*}%
Taking the reciprocal gives:
\begin{align*}
  \text{BF}_{01} &= \frac{1}{\text{BF}_{10}}\\
                 &= \frac{1}{0.6319}\\
  &=1.58 \;.
\end{align*}%
Assuming equal prior odds for $\mathcal{H}_0$ and $\mathcal{H}_1$, the posterior probability for $\mathcal{H}_0$ can be computed using Equation \ref{eq:probability} to be:%
\begin{align*}
  p(\mathcal{H}_0\mid \bm{y}) &= \frac{\text{BF}_{01}}{\text{BF}_{01}+1}\\
                              &= \frac{1.58}{1.58+1}\\
  &= 0.61240 \;.
\end{align*}%
Thus, data observed by \citet{fayol2012} are between 1.58 and 2.37 times more likely under $\mathcal{H}_0$ than under $\mathcal{H}_1$, with posterior probability for $\mathcal{H}_0$ between 0.61240 and 0.70326. Despite their claim for a null priming effect on multiplication, the evidence for this claim appears to be anecdotal at best.

\section{Simulation}
In this section, we describe a simulation study that we performed to benchmark the performance of the analytic Bayes factor in Proposition \ref{mainTheorem} against two other repeated-measures Bayes factors: the BIC approximation of \citet{faulkenberry2020} and the JZS Bayes factor of \citet{rouder2012}. In this simulation, we used randomly generated datasets that represented several different types of repeated-measures structure. Specifically, our datasets were generated from the linear mixed model%
\begin{equation}\label{simModel}
  y_{ij} = \mu + \alpha_j + \pi_i + \varepsilon_{ij};\hspace{5mm}i=1,\dots,n;\hspace{3mm}j=1,\dots,k,
\end{equation}%
where $\mu$ represents a grand mean, $a_j \sim \mathcal{N}(0,\sigma_{a})$ represent each of the $k$ randomly drawn treatment effects, $\pi_i \sim \mathcal{N}(0,\sigma^2_p)$ represent each of the $n$ randomly drawn participant effects, and $\varepsilon_{ij} \sim \mathcal{N}(0,\sigma_{\varepsilon}^2)$ represent the normally-distributed error terms. For convenience and brevity of exposition we set $k=3$, though we saw similar results with other values of $k$. Also, without loss of generality we set $\mu=0$ and $\sigma_{\varepsilon}=1$. We then systematically varied the following components of the model:

\begin{enumerate}
\item The number of experimental subjects $n$ was set to either $n=10$, $n=30$, or $n=80$;
\item The intraclass correlation $\rho$ among the subjects' repeated measurements was set to be either low ($\rho=0.2$) or high ($\rho=0.8$);
\item The size of the treatment effect was manipulated by setting $\tau = \sigma_a^2/\sigma^2$ to be either $\tau=0$, $\tau=0.5$, or $\tau=1$. For data generated under the condition $\tau=0$, the correct model is the null model $\h_0:\tau=0$, whereas for data generated under $\tau=0.5$ and $\tau=1.0$ the correct model is the alternative model $\h_1:\tau>0$.
    
\end{enumerate}

For each combination of number of subjects ($n=10,50,80$), treatment effect size ($\tau=0,0.5,1.0$), repeated-measures correlation ($\rho=0.2,0.8$), we generated 1000 simulated datasets. Each dataset was generated in R using the model in Equation \ref{simModel}. Each loop began with an empty vector which was then populated interatively in two nested loops. The outer loop was indexed over condition level $k=1,2,3$. The inner loop was indexed over subject number $n$. Then for each combination of condition level $i$ ($i=1,\dots,k=3$) and subject number $j$ ($j=1,\dots,n$), an observation was built additively as the sum of $\mu$ (which we assumed to be 0), a randomly drawn treatment effect $\alpha_i$, a randomly drawn subject effect $\pi_j$, and a randomly drawn noise term $\varepsilon_{ij}$. All randomly drawn components were assumed to be drawn from a normal distribution with mean $0$ and variance defined differently by term. For the treatment effect, the variance was assumed to be $\sigma^2_a = \tau\sigma^2$, and for the subject effect, the variance was assumed to be $\sigma^2_p = \sigma^2\rho/(1-\rho)$.

For each of the resulting datasets, we performed a repeated-measures analysis of variance, extracting the $F$ statistic and relevant degrees of freedom ($x$=between-treatments degrees of freedom and $y$=residual degrees of freedom). Then we used these values to compute two analytic Bayes factors from Proposition \ref{mainTheorem}: one using $\alpha=-\frac{1}{2}$ and another using $\alpha=0$. We also computed the BIC Bayes factor from \citet{faulkenberry2020}. Finally, we computed the JZS Bayes factor from \citet{rouder2012}; note that this Bayes factor can only be computed from the raw data, as the summary statistics alone are not sufficient for its estimation. However, as we wish to show that our analytic Bayes factor outperforms the BIC Bayes factor previously obtained in \citet{faulkenberry2020}, the JZS Bayes factor is an important target to assess against.

All obtained Bayes factors were converted to posterior probabilities via Equation \ref{eq:probability}, assuming 1-1 prior model odds. To compare the performance of the various computation methods in the simulation, we considered three analyses:

\begin{enumerate}
\item we visualized the distribution of posterior probabilities $p(\mathcal{H}_1\mid \bm{y})$;
\item we calculated the proportion of simulated trials for which the correct model was chosen (i.e., model choice accuracy);\item we calculated of the proportion of simulated trials for which both methods chose the same model (i.e., model choice consistency).
\end{enumerate}

\begin{figure}
  \centering
  \includegraphics[width=\textwidth]{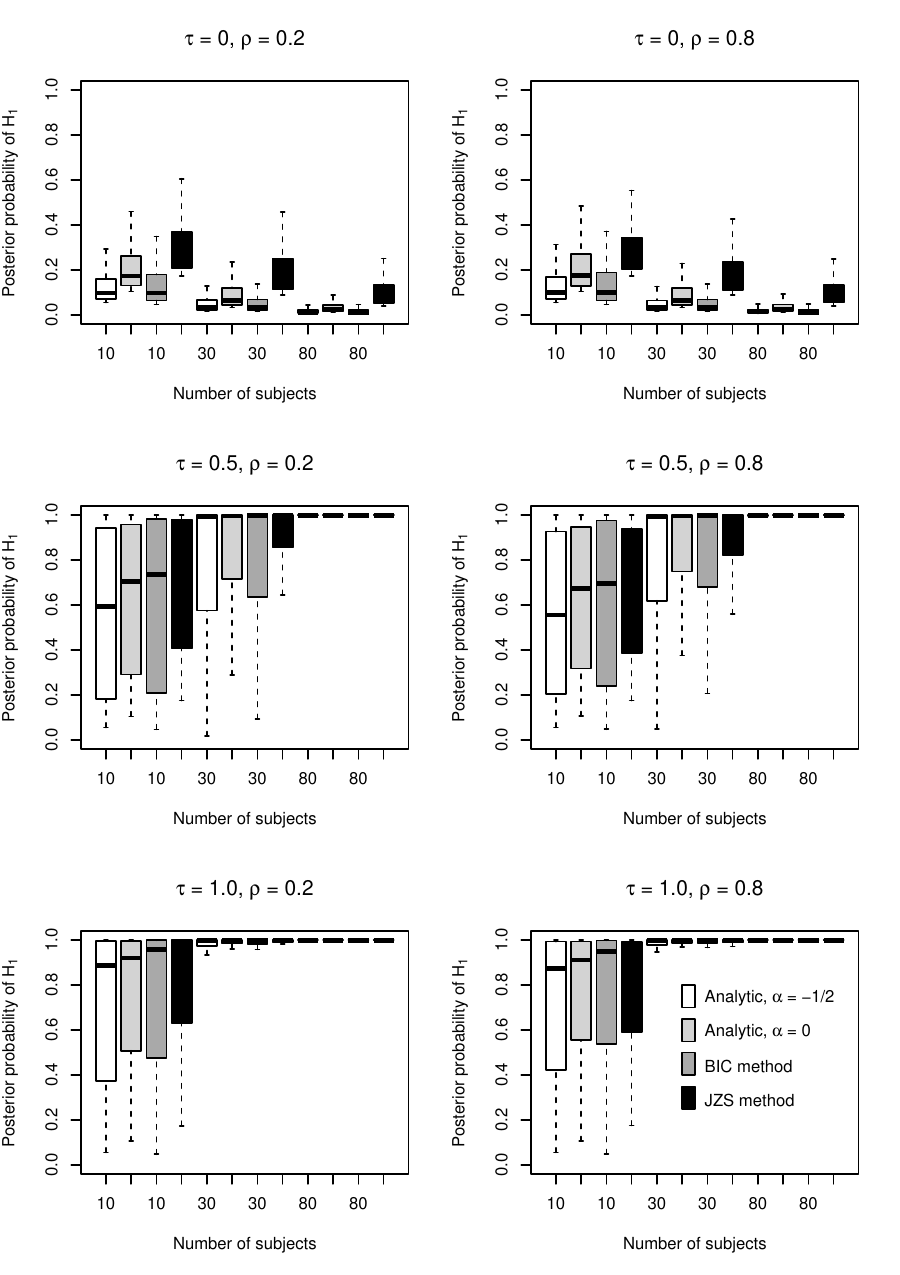}
  \caption{Boxplots depicting the distribution of the posterior probabilities $p(\mathcal{H}_0\mid \bm{y})$ for 1000 simulated datasets, split by number of subjects $n$ (within plots) and effect size $\tau$ and repeated-measures correlation $\rho$ (across plots). White and light-gray boxes represent the analytic Bayes factor with $\alpha=-\frac{1}{2}$ and $0$, respectively. Medium gray boxes represent the BIC Bayes factor of \citet{faulkenberry2020}. Black boxes represent the JZS Bayes factor of \citet{rouder2012}. }
  \label{fig:dist}
\end{figure}

First, let us consider the distribution of posterior probabilities $p(\mathcal{H}_1\mid \bm{y})$, displayed in Figure \ref{fig:dist}. Here, we constructed boxplots of the posterior probabilities for each of the four Bayes factor methods, split within plots by the number of subjects $n$, and split across plots to represent all possible combinations of effect size $\tau$ and repeated-measures correlation $\rho$. We can see that the variability of the posterior probability estimates decreases substantially as the number of subjects $n$ increases. For all simulated datasets in which $\h_0$ was the correct model (i.e., $\tau=0$, depicted in the first row of Figure \ref{fig:dist}), all methods produced posterior probabilities for $\h_1$ that were reasonably small. It is striking that in this case, the posterior probabilities derived from our new analytic Bayes factors as well as the BIC approximation of \citet{faulkenberry2020} are less than the posterior probabilities derived from the JZS Bayes factor. Note also that setting the prior width to $\alpha=-1/2$ produces the smallest posterior probabilities. We also note that the separation in performance between the four methods decreases with increasing numbers of subjects $n$.

For datasets in which $\h_1$ was the correct model (i.e., $\tau=0.5$ and $\tau=1.0$; rows 2 and 3 of Figure \ref{fig:dist}), a different pattern of results emerged. In these cases, the JZS Bayes factor produced posterior probabilities closer to 1 than did the analytic or BIC Bayes factors. Whereas setting $\alpha=-1/2$ was preferred in the $\tau=0$ case, these data reveal that $\alpha=0$ was the preferred setting when $\tau > 0$.  Finally, we note that repeated-measures correlation $\rho$ had little effect on the pattern of posterior probabilities that was observed.

For the next analysis, we calculated the proportion of simulated trials for which the correct model was chosen, the results of which are displayed in Table \ref{tab:acc}. This analysis is important because even though we see in Figure \ref{fig:dist} that the distributions of posterior probabilities appear to follow the same pattern across methods, it is unclear whether the analytic Bayes factor proposed in this paper provides the user with correct model choice. Since the data are simulated from target parameters, it is possible for us to gauge this performance exactly. For simulated datasets where $\tau=0$, the correct model is $\mathcal{H}_0$, whereas when $\tau=0.5$ or $\tau=1.0$, the correct model is $\mathcal{H}_1$. Thus, to compare performance of these Bayes factor methods, we calculated {\it model choice accuracy}, defined simply as the proportion of the 1000 simulated datasets for which the correct model ($\h_0$ or $\h_1$ was chosen. Model choice was defined by considering $\mathcal{H}_0$ to be chosen whenever $\text{BF}_{01}>1$ and $\mathcal{H}_1$ to be chosen whenever $\text{BF}_{01}<1$. 

\begin{table}
  \centering \small
  \begin{tabular}{cccccccccccc}
    & & \multicolumn{4}{c}{Correlation = 0.2} & & \multicolumn{4}{c}{Correlation = 0.8}\\
    & & $\alpha=-\frac{1}{2}$ & $\alpha=0$ & BIC & JZS & & $\alpha=-\frac{1}{2}$ & $\alpha=0$ & BIC & JZS\\
    \hline
    $\tau = 0$:\\
    $n=10$ & & .959 & .919 & .931 & .877 & & .959 & .924 & .928 & .894\\
    $n=30$ & & .980 & .966 & .976 & .920 & & .975 & .955 & .967 & .921\\
    $n=80$ & & .996 & .992 & .996 & .973 & & .990 & .981 & .989 & .965\\[2mm]

    $\tau = 0.5$:\\
    $n=10$ & & .554 & .626 & .612 & .682 & & .535 & .616 & .606 & .650\\
    $n=30$ & & .769 & .795 & .778 & .838 & & .782 & .817 & .788 & .858\\
    $n=80$ & & .892 & .910 & .897 & .932 & & .878 & .892 & .881 & .915 \\[2mm]

    $\tau = 1.0$:\\
    $n=10$ & & .700 & .756 & .746 & .813 & & .717 & .769 & .762 & .790\\
    $n=30$ & & .874 & .887 & .880 & .917 & & .861 & .886 & .869 & .904\\
    $n=80$ & & .927 & .937 & .931 & .948 & & .939 & .951 & .940 & .960\\
   \hline
    
  \end{tabular}
  \caption{Model choice accuracy calculated as the proportion of simulated datasets for which the correct model was chosen. Accuracies are presented as a function of Bayes factor method (analytic with $\alpha=-\frac{1}{2}$, analytic with $\alpha=0$, BIC, and JZS), numbers of subjects ($n=10,30,80$), effect size ($\tau=0,0.5,1.0$), and repeated-measures correlation ($\rho=0.2,0.8$).}
  \label{tab:acc}
\end{table}

Table \ref{tab:acc} shows that all methods were reasonably accurate at choosing the correct model. We observed the highest accuracy for the case where $\tau=0$; here, the analytic methods and the BIC method outperformed the JZS method (mirroring what we saw with the distributions of posterior probabilities in Figure \ref{fig:dist}). As before, setting $\alpha=-1/2$ for our analytic Bayes factor was the most accurate in this case. Accuracy when choosing $\h_1$ declined for all methods when $\tau>0$. Again, we see that when $\h_1$ is the correct model, the JZS Bayes factor is more accurate with respect to model selection. However, we note that setting $\alpha=0$ increased accuracy relative to $\alpha=-1/2$. Finally, we note that accuracy increased with increasing number of subjects $n$, and there was very little variation between our two repeated-measures correlation settings ($\rho=0.2$ and $\rho=0.8$). 

Finally, Table \ref{tab:con} displays model choice {\it consistency}, defined as the proportion of simulated datasets for which each of the methods based on summary statistics alone (analytic and BIC) chose the {\it same} model as the JZS Bayes factor. This analysis confirms what we observed in the previous two analyses: (1) consistency across methods increases with increasing numbers of subjects $n$; and (2) setting $\alpha=0$ gives Bayes factors that are more consistent with the JZS Bayes factor.

\begin{table}
  \centering \small
  \begin{tabular}{cccccccccc}
    & & \multicolumn{3}{c}{Correlation = 0.2} & & \multicolumn{3}{c}{Correlation = 0.8}\\
    & & $\alpha=-\frac{1}{2}$ & $\alpha=0$ & BIC  & & $\alpha=-\frac{1}{2}$ & $\alpha=0$ & BIC\\
    \hline
    $\tau = 0$:\\
    $n=10$ & & .918 & .958 & .946 & & .935 & .970 & .966\\
    $n=30$ & & .940 & .954 & .944 & & .946 & .966 & .954\\
    $n=80$ & & .977 & .981 & .977 & & .975 & .984 & .976\\[2mm]

    $\tau = 0.5$:\\
    $n=10$ & & .872 & .944 & .930 & & .885 & .966 & .956\\
    $n=30$ & & .931 & .957 & .940 & & .924 & .959 & .930\\
    $n=80$ & & .960 & .978 & .965 & & .963 & .977 & .966\\[2mm]

    $\tau = 1.0$:\\
    $n=10$ & & .887 & .943 & .933 & & .927 & .979 & .972\\
    $n=30$ & & .957 & .970 & .963 & & .957 & .982 & .965\\
    $n=80$ & & .979 & .989 & .983 & & .979 & .991 & .980\\                      
   \hline
    
  \end{tabular}
  \caption{Model choice consistency calculated as the proportion of simulated datasets for which each method chose the same model as the JZS Bayes factor. Proportions are presented as a function of Bayes factor method (analytic with $\alpha=-\frac{1}{2}$, analytic with $\alpha=0$, and BIC), numbers of subjects ($n=10,30,80$), effect size ($\tau=0,0.5,1.0$), and repeated-measures correlation ($\rho=0.2,0.8$).}
  \label{tab:con}
\end{table}

\section{Conclusion}

In this paper, we developed an analytic Bayes factor for repeated-measures analysis of variance designs. This Bayes factor gives researchers the ability to obtain Bayes factors directly from a minimal set of summary statistics (namely, the $F$ score and the degrees of freedom). This formula improves upon the repeated-measures BIC Bayes factor formula of \citet{faulkenberry2020} in two ways. First, this new analytic formula provides the user with an {\it exact} Bayes factor instead of an approximation. Second, our method gives the user the ability to tune the prior used in the computation of the Bayes factor by specifying a hyperparameter $\alpha$, which controls the width of the prior distribution of effect sizes $\tau=\sigma^2_a/\sigma^2$. Our simulation study shows that our analytic Bayes factor performs well compared to the JZS Bayes factor of \citet{rouder2012} (especially when the prior parameter $\alpha$ is set to 0). Remarkably, our analytic Bayes factor outperforms the JZS Bayes factor when data are generated under $\h_0$. We note that the analytic Bayes factor did not perform quite as well as the JZS Bayes on data generated under $\h_1$, but we think this limitation is far outweighed by ease of use of our method. First, our analytic Bayes factor has the unique ability to be computed {\it directly} from summary statistics, with no need for raw data or the need to compute a multi-dimensional integral.

We note that our simulation study indicates that model choice accuracy depends on the value of the prior parameter $\alpha$. Particularly, performance on datasets simulated under a null model $\h_0$ was better when we set $\alpha =-1/2$, whereas performance on datasets simulated under the alternative model $\h_1$ was better when $\alpha=0$. Thus, one recommendation for use of Proposition \ref{mainTheorem} in applied work would be to set $\alpha=-1/2$ when assessing evidence for a proposed null effect, and to set $\alpha=0$ when assessing evidence for a proposed non-null effect. Of course, an even better approach would be to use the entire consistency range of $\alpha$ and employ the ``bracketing'' procedure that we described in our examples above, thus providing a continuous range of Bayes factors rather than picking any one single end of that range.

In conclusion, we propose that our analytic Bayes factor will be an invaluable tool for anyone who wishes to assess the evidential value of their own data, as well as data from published studies where raw data is not readily available.

\begin{BLReferences}

\bibitem[Berger and Sellke, 1987]{berger1987}
Berger, J.~O. and Sellke, T. (1987).
\newblock Testing a point null hypothesis: {T}he irreconcilability of $p$
  values and evidence.
\newblock {\em Journal of the American Statistical Association}, 82(397):112.

\bibitem[Bortolussi and Dixon, 2002]{bortolussi2002}
Bortolussi, M. and Dixon, P. (2002).
\newblock {\em Psychonarratology: Foundations for the Empirical Study of
  Literary Response}.
\newblock Cambridge University Press.

\bibitem[Campbell and Thompson, 2012]{campbell2012}
Campbell, J. I.~D. and Thompson, V.~A. (2012).
\newblock {MorePower} 6.0 for {ANOVA} with relational confidence intervals and
  bayesian analysis.
\newblock {\em Behavior Research Methods}, 44(4):1255--1265.

\bibitem[Faulkenberry, 2018]{faulkenberry2018}
Faulkenberry, T.~J. (2018).
\newblock Computing {B}ayes factors to measure evidence from experiments: An
  extension of the {BIC} approximation.
\newblock {\em Biometrical Letters}, 55(1):31--43.

\bibitem[Faulkenberry, 2019]{faulkenberry2019}
Faulkenberry, T.~J. (2019).
\newblock Estimating evidential value from analysis of variance summaries: {A}
  comment on {L}y et al. (2018).
\newblock {\em Advances in Methods and Practices in Psychological Science}.

\bibitem[Faulkenberry, 2020]{faulkenberry2020}
Faulkenberry, T.~J. (2020).
\newblock Estimating {B}ayes factors from minimal summary statistics in
  repeated measures analysis of variance designs.
\newblock {\em Advances in Methodology and Statistics}, 17(1).

\bibitem[Faulkenberry, 2021]{faulkenberry2021}
Faulkenberry, T.~J. (2021).
\newblock The {P}earson {B}ayes factor: {A}n analytic formula for computing
  evidential value from minimal summary statistics.
\newblock {\em Biometrical Letters}, 58(1):1--26.

\bibitem[Fayol and Thevenot, 2012]{fayol2012}
Fayol, M. and Thevenot, C. (2012).
\newblock The use of procedural knowledge in simple addition and subtraction
  problems.
\newblock {\em Cognition}, 123(3):392--403.

\bibitem[Fisher, 1925]{fisher1925}
Fisher, R.~A. (1925).
\newblock {\em Statistical Methods for Research Workers}.
\newblock Oliver \& Boyd, Edinburgh.

\bibitem[García-Donato and Sun, 2007]{gds}
García-Donato, G. and Sun, D. (2007).
\newblock Objective priors for hypothesis testing in one-way random effects
  models.
\newblock {\em Canadian Journal of Statistics}, 35(2):303--320.

\bibitem[Gigerenzer, 2004]{gigerenzer2004}
Gigerenzer, G. (2004).
\newblock Mindless statistics.
\newblock {\em The {J}ournal of {S}ocio-{E}conomics}, 33(5):587--606.

\bibitem[Kass and Raftery, 1995]{kassRaftery1995}
Kass, R.~E. and Raftery, A.~E. (1995).
\newblock Bayes factors.
\newblock {\em Journal of the American Statistical Association},
  90(430):773--795.

\bibitem[Ly et~al., 2018]{ly2018}
Ly, A., Raj, A., Etz, A., Marsman, M., Gronau, Q.~F., and Wagenmakers, E.-J.
  (2018).
\newblock Bayesian reanalyses from summary statistics: {A} guide for academic
  consumers.
\newblock {\em Advances in Methods and Practices in Psychological Science},
  1(3):367--374.

\bibitem[Masson, 2011]{masson2011}
Masson, M. E.~J. (2011).
\newblock A tutorial on a practical {B}ayesian alternative to null-hypothesis
  significance testing.
\newblock {\em Behavior {R}esearch {M}ethods}, 43(3):679--690.

\bibitem[Nathoo and Masson, 2016]{nathoo2016}
Nathoo, F.~S. and Masson, M.~E. (2016).
\newblock Bayesian alternatives to null-hypothesis significance testing for
  repeated-measures designs.
\newblock {\em Journal of Mathematical Psychology}, 72:144--157.

\bibitem[Raftery, 1995]{raftery1995}
Raftery, A.~E. (1995).
\newblock Bayesian model selection in social research.
\newblock {\em Sociological {M}ethodology}, 25:111--163.

\bibitem[Rouder et~al., 2012]{rouder2012}
Rouder, J.~N., Morey, R.~D., Speckman, P.~L., and Province, J.~M. (2012).
\newblock Default {B}ayes factors for {ANOVA} designs.
\newblock {\em Journal of {M}athematical {P}sychology}, 56(5):356--374.

\bibitem[Schwarz, 1978]{schwarz1978}
Schwarz, G. (1978).
\newblock Estimating the dimension of a model.
\newblock {\em The Annals of Statistics}, 6(2):461--464.

\bibitem[Sellke et~al., 2001]{sellke2001}
Sellke, T., Bayarri, M.~J., and Berger, J.~O. (2001).
\newblock Calibration of $p$-values for testing precise null hypotheses.
\newblock {\em The American Statistician}, 55(1):62--71.

\bibitem[Thiele et~al., 2017]{thiele2017}
Thiele, J.~E., Haaf, J.~M., and Rouder, J.~N. (2017).
\newblock Is there variation across individuals in processing? {B}ayesian
  analysis for systems factorial technology.
\newblock {\em Journal of Mathematical Psychology}, 81:40--54.

\bibitem[Wagenmakers, 2007]{wagenmakers2007}
Wagenmakers, E.-J. (2007).
\newblock A practical solution to the pervasive problems of $p$ values.
\newblock {\em Psychonomic {B}ulletin {\&} {R}eview}, 14(5):779--804.

\bibitem[Wang and Sun, 2014]{wangSun}
Wang, M. and Sun, X. (2014).
\newblock Bayes factor consistency for one-way random effects model.
\newblock {\em Communications in Statistics - Theory and Methods},
  43(23):5072--5090.

\end{BLReferences}
\end{document}